*Mini Review*

# Quantitative Modeling of Action-Potential-Driven Sodium Dynamics in Neurons: From Sodium Influx to Sodium MRI

Yongxian Qian [1, 2, 3, *] OCRID: 0000-0002-9931-7790

[1] Bernard and Irene Schwartz Center for Biomedical Imaging, Department of Radiology, New York University Grossman School of Medicine, New York, New York 10016, USA.

[2] Center for Advanced Imaging Innovation and Research (CAI$^2$R), Department of Radiology, New York University Grossman School of Medicine, New York, New York 10016, USA.

[3] Neuroscience Institute, Department of Neuroscience and Physiology, New York University Grossman School of Medicine, New York, New York 10016, USA.

* **Corresponding author**. Email: Yongxian.Qian@nyulangone.org

**Conflict of interest:** YQ is one of the inventors of U.S. Patent Application: No.: 63/591,751. Title: *Process of multiple echo time acquisitions for quantitative bi-$T_2$ sodium magnetic resonance imaging*. Filed on October 19, 2023. IP Owner: New York University.

**Funding:** National Institutes of Health (NIH) RF1/R01 AG067502.

**Acknowledgments:** This work was performed under the rubric of the Center for Advanced Imaging Innovation and Research (CAI$^2$R, www.cai2r.net), an NIBIB National Center for Biomedical Imaging and Bioengineering (NIH P41 EB017183). The content is solely the responsibility of the author and does not necessarily represent the official views of the National Institutes of Health.

**Use of AI:** OpenAI's ChatGPT-6 was used to generate Figure 1 and to improve wording.

**Data Availability:** No data was generated in this work.

**Author Contribution:** YQ is the sole author responsible for the conceptualization, methodology, investigation, visualization, and writing (original draft, review, and editing) of this work.

**ABSTRACT:** Neuronal action potentials are accompanied by the influx of extracellular sodium ions ($Na^+$) into axons through voltage-gated sodium channels. Repetitive neuronal firing can lead to activity-dependent increases in intracellular sodium concentration, which are subsequently reversed primarily through the $Na^+/K^+$-ATPase. Quantifying these dynamic changes provides a biophysical foundation for investigating neuronal activity using sodium ($^{23}Na$) magnetic resonance imaging (MRI). This mini-review examines the relationships among firing frequency, action-potential-driven sodium influx, and intracellular sodium accumulation, to establish a quantitative framework linking neuronal activity to sodium MRI. The potential sensitivity and limitation of conventional and compartment-sensitive sodium MRI for detecting activity-dependent sodium changes are also discussed.

## 1. Introduction

Neuronal electrical activity (firing), manifested as action potentials (APs), is fundamental to information processing in the brain. APs propagate along axons through the coordinated movement of sodium and potassium ions across neuronal membranes. During the depolarization phase, voltage-gated sodium channels open, allowing extracellular $Na^+$ to flow into the axoplasm. This inward sodium ion current produces rapid depolarization of neuronal membrane that characterizes an action potential. Subsequently, the voltage-gated sodium channels become inactivation, but voltage-gated potassium channels are open, allowing intracellular $K^+$ to flow out of the cell and promoting membrane repolarization. The $Na^+/K^+$-ATPase then helps restore the transmembrane ionic gradients by transporting $Na^+$ out of the cell and $K^+$ into the cell, supporting the maintenance of ionic homeostasis and sustained neuronal activity.

Although the membrane potential changes by approximately 100 mV during an action potential, the amount of $Na^+$ entering the neuron during a single AP represents only a small fraction of total sodium present in the human brain tissue, for instance, ~0.4% in the white matter. Nevertheless, repetitive neuronal firing can lead to intracellular sodium accumulation. This change requires energy-dependent regulation, primarily through the $Na^+/K^+$-ATPase, to maintain and restore ionic homeostasis.

Quantifying the relationship between neuronal firing and sodium dynamics is important for understanding the physiological consequences of neuronal activity and developing noninvasive imaging approaches for investigating brain function. In particular, sodium MRI provides an opportunity to investigate tissue sodium concentration (TSC) and potentially characterize the relative contributions from intracellular and extracellular sodium compartments with appropriate acquisition and analysis techniques.

The central objective of this mini-review is to establish a quantitative framework linking neuronal firing frequency, action-potential-driven sodium influx, changes in intracellular sodium concentration, and the resulting sodium MRI signal. This review also examines how sodium redistribution between intracellular and extracellular compartments may influence MRI signal characteristics and discusses the potential and limitation of conventional and compartment-sensitive sodium MRI for investigating activity-dependent sodium dynamics.

## 2. Sodium influx during neuronal action potentials

### 2.1. Mechanism of sodium entry

At the resting membrane potential (Fig. 1), most voltage-gated sodium channels are closed but available for activation. When the membrane depolarizes to the action-potential threshold, these channels rapidly open, typically within a fraction of millisecond,[1,2] allowing $Na^+$ ions to enter the neuron down their electrochemical gradient, which is determined by both electrical and concentration gradients. The resulting inward sodium current further depolarizes the membrane, promoting the opening of additional voltage-gated sodium channels. This regenerative positive-feedback process produces the rapid rising phase of the action potential.[3,4] As depolarization progresses, voltage-gated sodium channels undergo rapid inactivation, typically on a timescale of approximately 1.0 ms after opening, reducing sodium influx and contributing to the termination of the depolarization phase.[2-4]

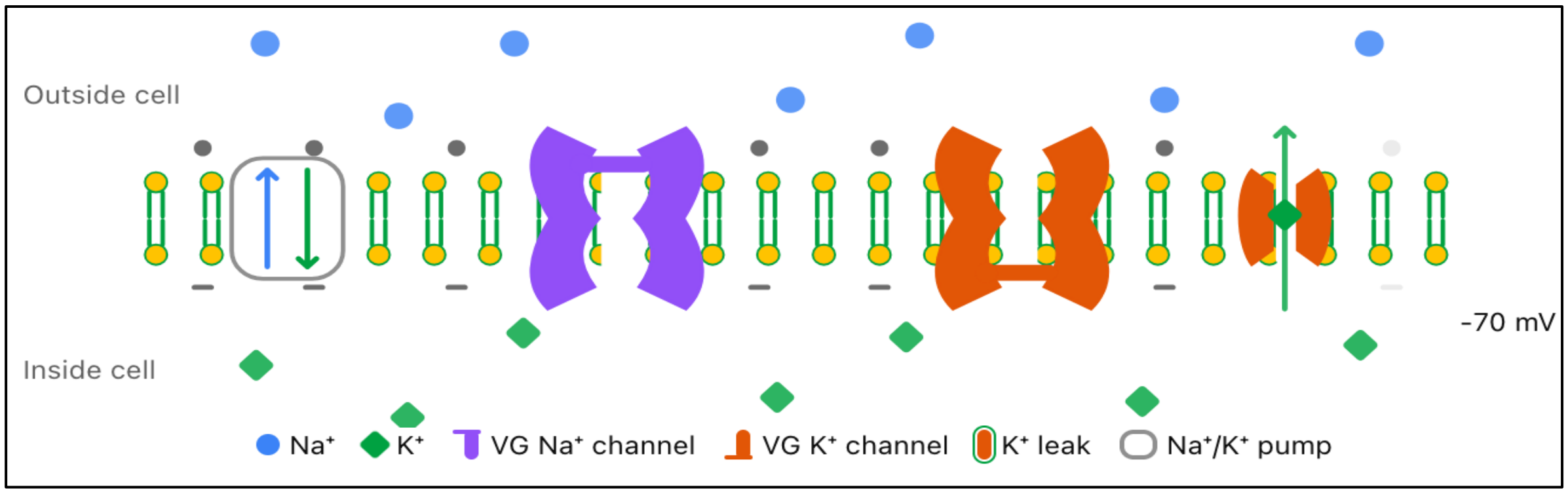

**Fig. 1.** Resting membrane potential. With voltage-gated (VG) channels closed, potassium leak helps set the resting voltage while the sodium-potassium pump maintains the ion gradients.

Meanwhile, voltage-gated potassium channels respond to membrane depolarization over a partially overlapping voltage range but activate more slowly than sodium channels. Their opening allows $K^+$ ions to flow out of the neuron down their electrochemical gradient. The resulting outward potassium current contributes to membrane repolarization, producing the falling phase of the action potential. Near the peak of action potential, sodium-channel inactivation and the increasing potassium conductance shift the balance of ionic currents from inward to outward, driving repolarization. Because potassium channels close gradually, potassium efflux continues as the membrane potential approaches to, and transiently falls below, its resting level, producing an afterhyperpolarization. As the membrane returns toward its resting potential, potassium channels progressively close, while sodium channels recover from inactivation and return to their closed, available state. The transmembrane sodium and potassium gradients are maintained and restored by ion-transport mechanisms, particularly the $Na^+/K^+$-ATPase, which transports three $Na^+$ ions out of the neuron in exchange for two $K^+$ ions entering the cell per ATP molecule hydrolyzed.[3,4]

The rapid activation and subsequent inactivation of voltage-gated sodium channels determine the time course of sodium entry during an individual action potential.[3] However, total sodium influx also depends on the number of channels activated, their conductance, and the duration of sodium current. These factors are important for quantifying sodium entry per action potential and estimating cumulative intracellular sodium load during repetitive neuronal firing.

Importantly, total sodium influx during an action potential is not determined solely by the amount of charge required to depolarize the neuronal membrane. Because sodium influx and potassium efflux can occur simultaneously during the action potential, the actual sodium influx per action potential, $Q_{\mathrm{Na}}$, can exceed the theoretical minimum, $Q_{\mathrm{min}}$, estimated from the membrane capacitance $C_{\mathrm{m}}$, membrane area A, and change in membrane potential ΔV, via $Q_{\mathrm{min}} = C_{\mathrm{m}}A\Delta V$.[3,4]

## 2.2. Experimental measurements of sodium influx

Carter and Bean (2009) experimentally investigated sodium entry during action potentials in mouse central neurons at physiological temperature (37°C).[3] By recording sodium currents elicited by individual action-potential waveforms and integrating these currents over the duration of each action potential, they quantified sodium influx per unit membrane area and revealed substantial difference among neuronal populations (Table 1). Specifically, sodium influx was associated primarily with action potential waveform, particularly spike duration and the extent of sodium-channel inactivation during repolarization. Fast-spiking neurons, including cerebellar Purkinje cells and cortical interneurons, exhibited greater sodium influx relative to the theoretical minimum required for membrane depolarization. The increased influx was attributed to incomplete sodium-channel inactivation during the falling phase of the action potential. The findings provide important experimental basis for estimating action-potential-driven sodium influx and modeling intracellular sodium accumulation under different neuronal firing conditions.

**Table 1.** Quantitative measurements of Na influx [3]

| **Neuronal population** | **$Na^+$ influx pmol/cm²/AP** |
|---|---|
| Cortical pyramidal neurons | 0.82 ± 0.30 |
| Cerebellar Purkinje neurons | 1.54 ± 0.54 |
| Cortical interneurons | 1.36 ± 0.27 |
| Hippocampal CA1 pyramidal neurons | 1.27 ± 0.65 |

## 2.3. Sodium influx as a function of firing frequency

Let $f$ represent neuronal firing frequency and let $q_{\mathrm{AP}}$ represent total sodium charge entering the neuron during one action potential. The number of sodium ions entering per AP is:

$$N_{AP} = q_{AP}/e \tag{1}$$

where $e$ is the elementary charge. For repetitive neuronal firing, the sodium influx rate can be expressed as:

$$J_{N_a} = fN_{AP} \tag{2}$$

Alternatively, using molar units:

$$\dot{n}_{N_a} = fq_{AP}/F \tag{3}$$

where $F$ is Faraday's constant.

For a neuron with multiple axonal branches, the total sodium influx depends on the cumulative membrane area activated by action potential propagation:

$$\dot{n}_{N_a,total} = \frac{f}{F}\sum_{k=1}^{m} q_{AP,k} \tag{4}$$

These formulations provide the first component of the proposed quantitative model: neuronal firing frequency determines the rate of sodium entry, subject to the sodium influx associated with each action potential.

# 3. Conversion of sodium influx into intracellular sodium concentration changes

The amount of sodium entering a neuron does not directly determine its intracellular concentration change. The latter also depends on the volume of the cellular compartment receiving the sodium ions. For a cylindrical axon with radius $r$ and length $L$, the membrane surface area and intracellular volume are $A_m = 2\pi rL$ and $V_i = \pi r^2 L$, respectively.

Let $q_{Na}$ represent sodium influx per unit membrane area per action potential, expressed in mol/cm²/AP. The increase in intracellular sodium concentration, $\Delta C_{i,AP}$, following one action potential is:

$$\Delta C_{i,AP} = \frac{q_{Na} A_m}{V_i} \tag{5}$$

Substituting the cylindrical geometry:

$$\Delta C_{i,AP} = \frac{2q_{Na}}{r} \tag{6}$$

This relationship demonstrates that smaller axons experience greater intracellular sodium concentration changes per action potential for the same sodium influx per unit membrane area, due to their larger membrane surface-area-to-volume ratio.

### 3.1. Numerical estimation

Consider a representative unmyelinated axon with the following parameters: diameter, 0.3 µm; length, 1.0 cm; and sodium influx, 0.82 pmol/cm²/AP. Using the experimental value reported for the cortical pyramidal neurons:

$$q_{N_a} = 0.82 \times 10^{-12} \text{ mol/cm}^2\text{/AP}$$

The predicted intracellular sodium concentration increment is:

$$\Delta C_{i,AP} = \frac{2(0.82\times10^{-12})}{1.5\times10^{-5}} \approx 0.109 \text{ (mM/AP)}$$

Thus, approximately 0.109 mM of sodium would accumulate within this axonal segment following one action potential if all entering sodium remained within the segment.

### 3.2. Dependence on neuronal firing frequency

If sodium removal is initially neglected, repetitive action potentials produce cumulative sodium accumulation according to:

$$\Delta C_{i,AP}(t) = f\Delta C_{i,AP}\,\Delta t \tag{7}$$

Using the calculated value of 0.109 mM/AP, the theoretical gross sodium accumulation rates at typical neuronal firing frequencies in the human brain are illustrated in Table 2.

These values represent gross sodium influx rather than sustained intracellular accumulation. In physiological conditions, sodium extrusion, diffusion, and changes in neuronal excitability continuously influence the actual concentration.

**Table 2.** Estimated gross sodium accumulation with firing frequency

| Firing frequency, Hz | Gross accumulation rate, mM/s | Firing frequency, Hz | Gross accumulation rate, mM/s | Firing frequency, Hz | Gross accumulation rate, mM/s |
|---|---|---|---|---|---|
| 1 | 0.109 | 20 | 2.18 | 150 | 16.35 |
| 5 | 0.545 | 50 | 5.45 | 200 | 21.8 |
| 10 | 1.09 | 100 | 10.9 | 300 | 32.7 |

## 4. Dynamic regulation of intracellular sodium concentration

### 4.1. Sodium extrusion through the $Na^+/K^+$-ATPase

Following neuronal firing, intracellular sodium is gradually removed by the $Na^+/K^+$-ATPase. Each pump cycle transports three $Na^+$ ions out of the cell while importing two $K^+$ ions, consuming one adenosine triphosphate (ATP) molecule. Consequently, intracellular sodium concentration is determined by the balance between sodium influx and active extrusion. A simplified first-order model can describe this relationship:

$$\frac{dC_i}{dt} = f\Delta C_{i,AP} - \frac{C_i - C_{i,0}}{\tau_{Na}} \tag{8}$$

where $C_i$ is intracellular sodium concentration at time $t$, $C_{i,0}$ is baseline intracellular sodium concentration at $t$=0, and $\tau_{Na}$ is the effective sodium recovery time constant. The first term describes sodium influx associated with neuronal firing, whereas the second term approximates sodium removal through homeostatic mechanisms.

For constant firing frequency, the solution is:

$$C_i(t) - C_{i,0} = f\Delta C_{i,AP}\tau_{Na}(1 - e^{-t/\tau_{Na}}) \tag{9}$$

At steady state ($t$=∞):

$$\Delta C_{i,steady} \equiv C_i(\infty) - C_{i,0} = f\Delta C_{i,AP}\tau_{Na} \tag{10}$$

Equation (10) provides a simple quantitative relationship between neuronal firing frequency and intracellular sodium concentration. It predicts that increasing neuronal firing frequency produces greater intracellular sodium accumulation, while a longer sodium recovery time increases the magnitude of the accumulation. However, this first-order approximation assumes constant firing-related sodium influx and linear sodium recovery. Actual $Na^+/K^+$-ATPase activity depends on intracellular sodium concentration, extracellular potassium concentration, ATP availability, and other physiological factors.

## 4.2. A more physiologically realistic model

A more advanced model should incorporate nonlinear sodium pump kinetics. One possible formulation is:

$$P_{Na} = P_{Na,max} \times \frac{C_i^n}{K_{Na}^n + C_i^n} \times \frac{C_{e,K}^m}{K_K^m + C_{e,K}^m} \tag{11}$$

where: $P_{Na}$ is the sodium extrusion rate, $P_{Na,max}$ is the maximum pump-mediated sodium extrusion rate, $K_{Na}$ and $K_K$ are effective concentration constants for sodium and potassium ions, respectively, $n$ and $m$ are empirical Hill coefficients, and $C_{e,K}$ represents extracellular potassium concentration.[5]

The intracellular sodium equation can then be written as:

$$\frac{dC_i}{dt} = f\Delta C_{i,AP} - P_{Na}(C_i, C_{e,K}) \tag{12}$$

This nonlinear formulation would provide a more realistic description of intracellular sodium dynamics during prolonged or high-frequency neuronal stimulation. Cressman et al. (2009) [6] provide a useful foundation for this type of modeling through their neuronal model incorporating sodium and potassium concentration dynamics, membrane excitability, and sodium pump activity.

## 5. Intracellular and extracellular sodium redistribution

### 5.1. Two-compartment sodium model

An essential consideration for sodium MRI is that neuronal firing redistributes sodium between intracellular and extracellular compartments. A simplified two-compartment model can be established:

$$C_{TSC} = a_i C_i + a_e C_e \text{ , with } a_i + a_e = 1 \quad (13)$$

where $C_{TSC}$ is total tissue sodium concentration, and $a_i$ and $a_e$ are intracellular and extracellular sodium volume fraction, respectively.

Representative sodium concentrations in the human brain tissue are approximately 15 mM intracellularly and 145 mM extracellularly.[7] However, these concentrations and the volume fractions of the intracellular and extracellular compartments vary across brain regions and tissue types.[8] For example, the extracellular volume fraction is typically ~20% in the gray matter and ~10% in the white matter, corresponding to intracellular volume fractions of ~80% and ~90%, respectively.[7,9-11] Using these representative values, the volume-weighted total tissue sodium concentration, $C_{TSC}$, can be estimated as 41 mM in the gray matter and 28 mM in the white matter. These values provide useful representative estimates for modeling sodium distribution between intracellular and extracellular compartments, although actual values depend on tissue composition and physiological conditions.

In vivo, sodium redistribution is also influenced by extracellular diffusion, astrocytic sodium uptake, vascular exchange, and cellular volume changes.

### 5.2. A critical implication for sodium MRI

During neuronal firing, sodium ions entering neurons from the extracellular space can increase intracellular sodium concentration. However, the extracellular sodium concentration may remain approximately constant because the extracellular space is continuously connected to surrounding tissue in the brain, allowing sodium ions to diffuse across imaging voxel boundaries and replenish local extracellular sodium. Under this open-compartment assumption, intracellular sodium accumulation can increase the total tissue sodium concentration within an imaging voxel, even when extracellular sodium concentration remains relatively stable. This provides a physiological rationale for using conventional TSC measurements to investigate activity-dependent changes in intracellular sodium concentration.[12]

Nevertheless, the sensitivity of conventional sodium MRI to these changes may be limited by the substantially higher extracellular sodium concentration (145 mM) compared with intracellular sodium concentration (15 mM). The relatively large extracellular sodium contribution can dominate the total sodium MRI signal, potentially obscuring smaller activity-dependent changes in intracellular sodium. This limitation highlights the potential importance of compartment-sensitive sodium MRI techniques [7,15-17] for investigating neuronal sodium dynamics.

## 6. Modeling activity-dependent sodium MRI signals

### 6.1. Total sodium MRI signal

The measured sodium MRI signal depends on sodium concentration, relaxation properties, acquisition parameters, and compartment-specific signal contributions. A simplified two-compartment signal model is:

$$S_{Na} = g(a_i C_i X_i + a_e C_e X_e) \tag{14}$$

where $g$ is an MRI system-dependent scaling factor, $X_i$ and $X_e$ represent the signal-weighting factors for intracellular and extracellular sodium, respectively. These factors depend on relaxation characteristics, including $T_1$, $T_2$, and $T_2$*, and acquisition parameters, including echo time (TE), repetition time (TR), and RF excitation flip angle.

For a more detailed model, the signal contributions can be represented using relaxation-weighted terms:

$$X_j = (1 - e^{-TR/T_{1,j}})e^{-TE/T^*_{2,j}} \text{ , with } j = i, e \tag{15}$$

This expression assumes an appropriate excitation and a mono-exponential $T_2^*$ decay in sodium signal under the condition of relatively fast motion: $\omega_0 << 1/\tau_c$, i.e., the Larmor frequency $\omega_0$ is much smaller than the inverse of the correlation time $\tau_c$ of sodium ions interacting with surrounding electric field gradient.[13]

When the fast motion condition is not satisfied, sodium signals decay biexponentially because sodium nuclei have spin 3/2 and their quadrupolar interactions with surrounding electric field gradients produce complex relaxation behavior. Sodium signals often exhibit biexponential decay in $T_2$ relaxation, with a short relaxation time $T_{2,s}$ of about 1–5 ms and a long relaxation time $T_{2,l}$ of about 15–35 ms in the human brain tissue, for instance.[7] Consequently, the single-$T_2^*$ relaxation in Eq. (15) is replaced with biexponential relaxation,

$$Y_j = b_s e^{-TE/T^*_{2,s,j}} + b_l e^{-TE/T^*_{2,l,j}} \text{, with } j = i, e \tag{16}$$

where $b_s$ and $b_l$ represent relative signal intensity of individual sodium nuclear spins, with theoretical values of 0.6 and 0.4, respectively.[13]

The biexponential relaxation can exist in both intracellular and extracellular compartments, but mainly in the intracellular compartment in the human brain tissue due to relatively slow motion and a large population of negatively charged proteins and macromolecules. The short-$T_2$ relaxation is very fast, making ultrashort echo time (UTE) acquisitions [14] important for detecting the rapidly decaying signal component.

### 6.2. Compartment-sensitive sodium MRI

Several approaches have been developed to highlight intracellular sodium contributions, including triple-quantum-filtered (TQF) sodium MRI,[15] double single-quantum (DSQ) acquisitions,[16] inversion-recovery (IR) sodium MRI,[17] biexponential relaxation-weighted imaging.[7] These approaches exploit differences in sodium relaxation and molecular interactions between compartments. However, intracellular sodium is not necessarily synonymous with biexponential-$T_2$ sodium, nor is extracellular sodium necessarily equivalent to mono-$T_2$ sodium. Thus, compartment-sensitive MRI measurements require appropriate calibration and biophysical modeling. Egidi et al. (2025) provide a recent review of these approaches and their limitations in human brain sodium MRI.[16]

## 7. Proposed integrated quantitative model

Based on the considerations reviewed above, an integrated model should connect neuronal firing frequency, sodium influx, sodium extrusion, and MRI signal formation. The proposed model consists of two principal equations. The first one is Eq. (12), describing intracellular sodium dynamics, while the second one is Eq. (14), representing sodium MRI signal. Together, these equations provide a framework for predicting how changes in neuronal firing frequency may influence intracellular sodium concentration and the measured sodium MRI signal (Figure 2).

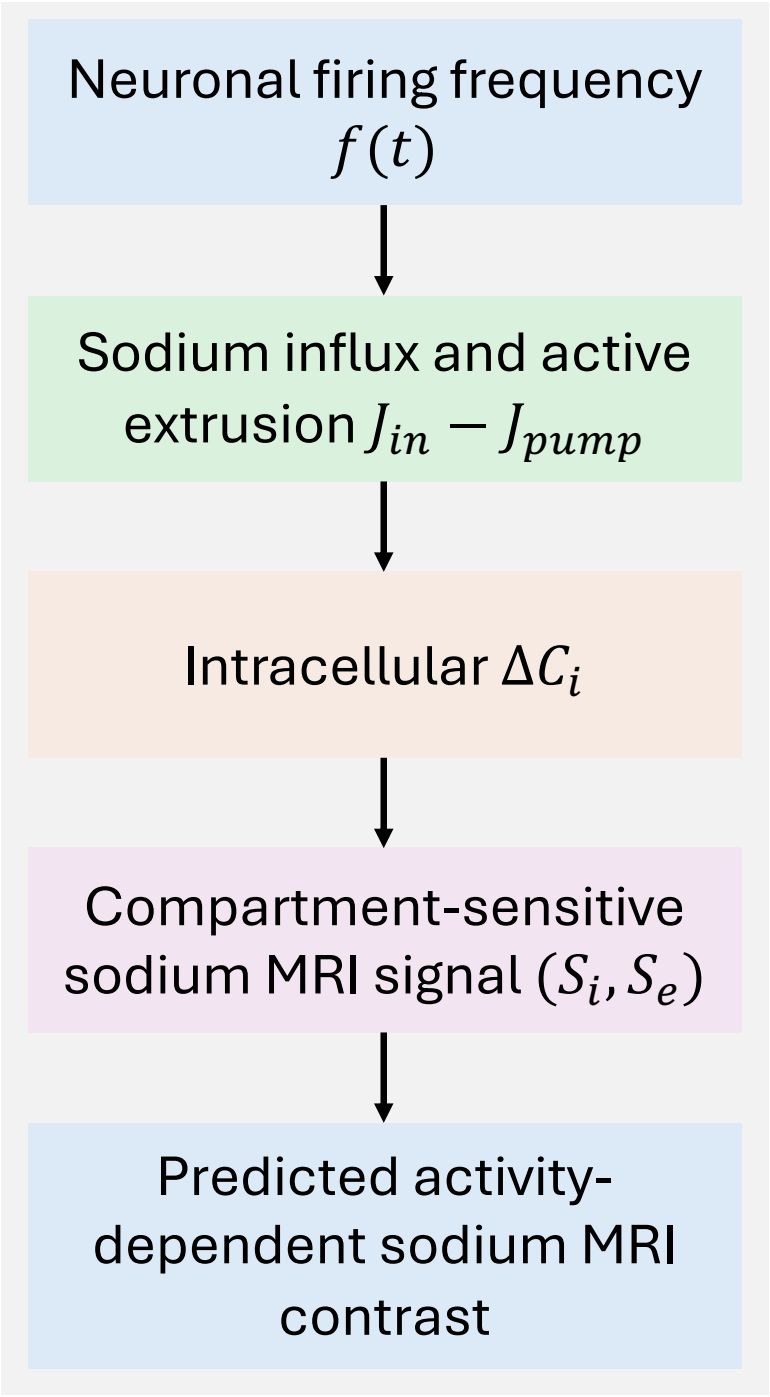

**Fig. 2.** Proposed model architecture

This model can initially be implemented as a system of ordinary differential equations for a single neuronal compartment. Spatial diffusion and multiple cell types can be incorporated in subsequent stages.

## 8. Experimental validation and model development

The proposed model requires experimental validation to establish whether predicted sodium concentration changes are physiologically realistic and detectable using sodium MRI. Three complementary experimental approaches are recommended.

### 8.1. Electrophysiological measurements

Electrophysiological recordings should establish the relationship between stimulation conditions and neuronal firing characteristics. Suggested stimulation frequencies include 1, 5, 10, and 20 Hz, as well as higher frequencies listed in Table 2. For each condition, action potential frequency, waveform, and duration should be recorded. Importantly, the integrated sodium current per action potential should be measured or estimated for the neuronal population under investigation rather than assumed to be constant.

### 8.2. Sodium-sensitive fluorescence imaging

Quantitative sodium-sensitive fluorescence imaging can independently measure intracellular sodium concentration changes during neuronal stimulation. Gerkau et al. (2019) [19] investigated activity-induced intracellular sodium transients and ATP dynamics in mouse hippocampal neurons. Their results demonstrated that global neuronal sodium influx can produce measurable changes in intracellular ATP, illustrating the close relationship between sodium homeostasis and neuronal energy metabolism. These measurements could be used to validate the predicted sodium accumulation and recovery time constants.

### 8.3. Sodium MRI measurements

Finally, $^{23}$Na-MRI can be used to investigate whether the predicted sodium redistribution produces measurable MRI signal changes. A suitable experimental design would compare baseline, neuronal stimulation, and post-stimulation recovery conditions. Both conventional tissue sodium concentration measurements and compartment-sensitive acquisitions should be considered. The principal outcome measures would include: (1) the changes in total tissue sodium concentration, (2) the changes in the bi-$T_2$ sodium concentration or signal fraction that is associated with intracellular sodium, and (3) temporal recovery of sodium signals following stimulation.

The combination of electrophysiology, sodium-sensitive fluorescence imaging, and sodium MRI would provide a comprehensive approach to evaluating the proposed model.

## 9. Potential applications and scientific significance

The quantitative framework may have several important applications in neuroscience and neuroimaging. First, it could provide a theoretical relationship between neuronal firing frequency and intracellular sodium concentration, allowing the magnitude of activity-dependent sodium changes to be estimated. Second, it could help determine the temporal characteristics of sodium accumulation and recovery following neuronal stimulation, which are important for understanding the temporal sensitivity of sodium MRI. Third, it could provide a basis for investigating compartment-sensitive sodium MRI as a potential method for studying neuronal function.

An important distinction should be maintained between measuring neuronal electrical activity and measuring the downstream ionic consequences of neuronal activity. Sodium MRI does not directly measure individual action potentials. Rather, it may detect changes in sodium concentration or compartmentalization associated with neuronal activity. This distinction is particularly important when evaluating sodium MRI as a complementary technique to the blood-oxygenation-dependent (BOLD) functional MRI.[20,21]

## 10. Conclusion

Neuronal action potentials produce sodium influx through voltage-gated sodium channels, resulting in transient changes in intracellular sodium concentration. Repetitive neuronal firing can produce cumulative sodium accumulation, which is subsequently reversed through the $Na^+/K^+$-ATPase and other ionic homeostatic mechanisms.

A quantitative model incorporating sodium influx, neuronal geometry, firing frequency, sodium extrusion, and intracellular concentration change provides a theoretical framework for predicting activity-dependent sodium concentration changes. Integrating this framework with compartment-specific sodium MRI signal modeling may help determine whether these changes are detectable using conventional or compartment-sensitive sodium MRI.

Further experimental validation is required to establish the sensitivity, temporal characteristics, and physiological specificity of sodium MRI for investigating neuronal activity.